%% file: main.tex
\documentclass[conference]{IEEEtran}
\IEEEoverridecommandlockouts

\usepackage{cite}
\usepackage{amsmath,amssymb,amsfonts,amsthm}
\usepackage{graphicx}
\usepackage{textcomp}
\usepackage{xcolor}
\usepackage{bm}
\usepackage{accents}
\usepackage{physics}
\usepackage{soul}
\usepackage{asymptote}
\usepackage{tikz}
\usetikzlibrary{calc,backgrounds}
\usepackage{tikz,pgfplots}
\usetikzlibrary{positioning, shapes, arrows.meta}
\pgfplotsset{compat=1.18}
\usepackage[caption=false,font=footnotesize]{subfig}

\usepackage{algpseudocode}
\usepackage[linesnumbered,ruled,vlined]{algorithm2e}
\SetKwInput{KwInput}{Inputs}                
\SetKwInput{KwOutput}{Outputs}
\SetKwInput{KwInit}{Initialization}

\newcommand{\nris}{N}
\newcommand{\ny}{N_{y}}
\newcommand{\nx}{N_{x}}
\newcommand{\sn}{\mathbf{s}_{n}}

\newcommand{\UEk}[1]{\mathbf{u}_{#1}}

\newcommand{\ori}{\mathbf{g}}
\newcommand{\evec}{\mb{e}}

\newcommand{\dy}{{\rm d}_y}
\newcommand{\dx}{{\rm d}_x}
\newcommand{\du}{{\rm d}_u}
\newcommand{\pinv}[1]{#1^{\dag}} 
\newcommand{\mb}[1]{\mathbf{#1}} 

\def\BibTeX{{\rm B\kern-.05em{\sc i\kern-.025em b}\kern-.08em
    T\kern-.1667em\lower.7ex\hbox{E}\kern-.125emX}}
\begin{document}

\title{Near-field Anchor-free Localization using Reconfigurable Intelligent Surfaces
}

\author{Srikar Sharma Sadhu,  Praful D. Mankar, and Santosh Nannuru
\thanks{S. S. Sadhu,  P. D. Mankar and S. Nannuru are with Signal Processing and Communication Research Center, IIIT Hyderabad, India. Email: srikar.sadhu@research.iiit.ac.in, \{praful.mankar,santosh.nannuru\}@iiit.ac.in.  
} 
}

\maketitle

\input{0_Abstract}

\section{Introduction}\label{sec:Introduction}
\input{1_Intro}

\section{System Model}\label{sec:system_model}
\input{2_System_Model}

\section{RIS-assisted Anchor-free Localization}\label{sec:phase_design}
\input{3_RIS_assisted_anchor_free}

\section{Numerical Results and Discussion} \label{sec:results}
\input{4_Numerical_Results}

\section{Conclusion}
\input{5_Conclusion}

\bibliographystyle{IEEEtran}  
\input{main.bbl}

\end{document}

%% file: 0_Abstract.tex
\begin{abstract}
Near-field localization is expected to play a crucial role in enabling a plethora of applications under the paradigm of 6G networks. The conventional localization methods rely on complex infrastructure for providing cooperative anchor nodes that often contribute to higher network overload and energy consumption. To address this, the passive reconfigurable intelligent surfaces (RISs) can be leveraged as perfectly synced reference nodes for developing anchor-free near-field localization. First, we obtain the optimal RIS configurations that maximizes the block-wise averaged trace of Fisher information matrix so that localization error variance can be minimized across the area-of-interest (AoI). Next, we present a two-stage anchor-free localization framework wherein first a coarse estimate is obtained using cosine similarity between the coarse grid and the signal received under pre-defined optimal RIS configurations. In second stage, we refine solution similarly using a finer grid constructed around the coarse estimate. The numerical results show that the proposed RIS-aided anchor-free localization provides small root mean square error for practical values of signal-to-noise ratio (SNR), RIS dimension, and number of antennas at user.
\end{abstract}
\begin{IEEEkeywords}
RIS, near-field localization, anchor-free localization. 
\end{IEEEkeywords}

%% file: 1_Intro.tex
Radio localization is emerging as an alternative to GPS-based localization, especially in environments like indoor scenario where wireless networks can be  accessed more reliably as compared to satellite signals  \cite{Kamran_2023}. Localization is expected to play an important role in the  integration of cellular communication with a plethora of new applications  emerging under 6G paradigm such as sensing, object mapping, massive IoT, etc., \cite{Zhao_6g_white_paper_near_field,bourdoux20206gwhitepaperlocalization}. 
However, the conventional localization approaches rely on  multiple active anchors such as base stations (BSs) and access points (APs) that often have high hardware complexity  and  energy consumption, and requires tight synchronization \cite{srikar_2025,Sarah_2021,1gto5g_2018}. 
Interestingly, the advent of  
reconfigurable intelligent surfaces (RISs) offers an attractive solution to this problem where the passive RISs can be utilized  as the tightly synced reference points for localization.  In particular, the  signals reflected off the passive RIS surfaces  can be utilized for triangulation and estimating the angle of arrival (AoA), angle of departure (AoD), time difference of arrival etc. without explicitly requiring  synchronization. Motivated by this, we focus on utilizing RIS for realizing {\em anchor-free localization}.

\subsection{Related Works}
Conventional radio localization wherein an agent tries to localize itself using a set of active anchors has been explored thoroughly in literature for a variety of system settings. Interested readers may refer to \cite{Cheng_2012_survey,1gto5g_2018,Zafari_2019} for a comprehensive survey along these lines. Additionally, the authors of \cite{Tapas_2018,Sekhar_2021,Premachandra_2023,Cai_2024}  present anchor-free localization which involves the user equipment (UE) localizing itself in the absence of  BSs or APs. The authors of \cite{Tapas_2018,Sekhar_2021} use a cooperative environment for localizing the UEs via assistance of  a special node that is either GPS-enabled or acts as a central hub.
Alternatively, the authors of \cite{Premachandra_2023,Cai_2024} rely on the ultra-wideband protocol in anchor-free enviornments. In particular, \cite{Premachandra_2023} presents trilateration based simultaneous localization and mapping (SLAM), whereas \cite{Cai_2024} utilizes peer-to-peer range measurements to perform 3D pose estimation.
However, these works still require a certain level of cooperation. This either requires the infrastructural dependencies for tagging at least one node or peer-to-peer communication, which essentially limits these works from truly being anchor-free localization.

\begin{figure*}[ht]
    \centering
    \subfloat[Near-field area of interest \label{fig:aoi}]{\includegraphics[]{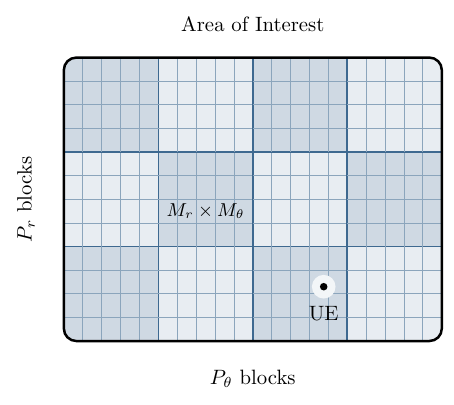}}
    \hspace{3cm}
    \subfloat[System model\label{fig:system}]{\includegraphics[]{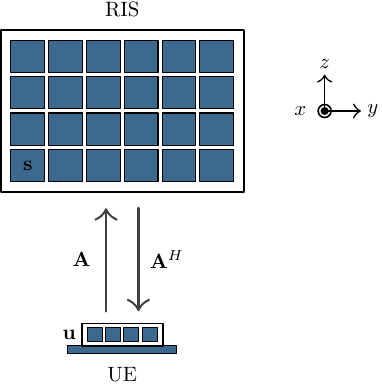}}
    \caption{An illustration of RIS-assisted near-field anchor-free localization.}
    \label{fig:system_model}
\end{figure*}

In contrast, the RIS can facilitate localization without requiring the system level dependence \cite{Chen_2023} where the UE can use the  signal reflected from  RIS  to localize itself with respect to RIS position --- even the UE location based dynamic RIS control is not necessary (as will be shown in this paper). 
However, this line of research remains to be explored in the literature to its fullest potential. For example,  \cite{Kamran_2022,Kim_2024,Mustafa_2025,Kim_2025_FMCW} are the only  available works on RIS-assisted anchor-free localization.
The authors of  \cite{Kamran_2022,Kim_2024,Mustafa_2025} developed   frameworks wherein they utilize the prior knowledge of the UE location to optimally configure the RIS and then employ two-stage approach involving coarse estimation and its refinement via iterative algorithms under optimally configured RIS.
In particular, the authors of \cite{Kamran_2022} obtain the coarse estimate using delay profile that is obtained via the inverse fast Fourier transform of the received signal. Next, the authors   refine the coarse estimate by maximizing the likelihood function.
In \cite{Kim_2024}, the authors develop anchor-free SLAM wherein the RIS phase shifts require to be dynamically configured based on the UE state. The  UE state and landmarks are estimated using a modified marginal Poisson multi-Bernoulli SLAM filter.
While \cite{Kamran_2022,Kim_2024} use full duplex UE for RIS-assisted localization/SLAM, the authors of \cite{Mustafa_2025}  consider half-duplex UE localization with cooperation from multiple other UEs for receiving the signal reflected from RIS to facilitate the maximum likelihood estimation of the UE's location. On the other hand, the authors of \cite{Kim_2025_FMCW} consider the  reception of signal reflected from RIS  at full-duplex  UE under a phase shift codebook that  sweeps the entire angular region.  This facilitates the UE to  construct a delay-Doppler map corresponding to the non-stationery objects in the environment, which is further used to localize the UE.
Although, these works present novel approaches, the RIS phase shift needs to be dynamically controlled depending  on the UE location \cite{Kamran_2022,Kim_2024,Mustafa_2025}. This requires frequent adaptation and leads to increased overhead and computational complexity.

Motivated by this, we present the design of RIS configurations that maximize the block-wise averaged trace of Fisher information matrix (FIM) in order to lower the localization error variance across an area-of-interest (AoI). This mitigates the need for frequent RIS control, as opposed to the previous works.  
We developed an RIS-assisted near-field anchor-free localization method that first obtains multiple coarse estimates of UE location by evaluating the cosine similarity between a coarse grid constructed over the AoI and the signal received under the optimal RIS configurations.
Next, for further refinement, we employ a similar approach using a fine-resolution grid constructed around the coarse estimates.  Through extensive numerical analysis, we show that the proposed approach provides low root mean square error (RMSE) performance for practical values of signal-to-noise ratio (SNR). We further show the impact of grid resolution, RIS dimension and number of UE antennas on the RMSE performance.

\subsection{Notation}
Vectors are denoted using bold lowercase letters (e.g. $\mb{q}$) and matrices are denoted using bold uppercase letters (e.g. $\mb{Q}$).
Transpose, conjugate, conjugate transpose and pseudo-inverse are represented as $(\cdot)^T$, $(\cdot)^*$, $(\cdot)^H$ and $\pinv{(\cdot)}$, respectively.
The operators $\operatorname{tr}(\cdot)$, $\operatorname{diag}(\mb{q})$ and $\operatorname{vec}\left(\cdot\right)$ represent the trace of a matrix, a diagonal matrix with diagonal $\mb{q}$ and the vectorized version of a matrix respectively.
$\mb{I}_Q$ represents a $Q \times Q$ identity matrix. 
Vector and Frobenius norms are denoted using $\norm{\cdot}$ and $\norm{\cdot}_F$ respectively.

%% file: 2_System_Model.tex
This work considers anchor-free localization of a UE
assumed to be equipped with a uniform linear array (ULA) of length $K$ lying in a 2D AoI denoted as $\mathcal{A}$.
The AoI (see Fig. \ref{fig:aoi}) comprises $P = P_r \times P_{\theta}$ blocks, each of size $M = M_r \times M_{\theta}$ such that $\mathcal{A}$ contains $P \times M$ grid points. The AoI as illustrated in Fig. \ref{fig:aoi} considered to be the near-field region of an RIS.
The UE is located off-grid at $\mb{u}$ in the $xy$-plane at $z=0$, with ULA oriented along the $y$ axis, as shown in Fig. \ref{fig:system}.
Here, $\mb{u}=r\evec$ such that $r$ is distance from origin and $\evec \triangleq \left[\cos\theta,~\sin\theta,~0\right]^T$ represents the unit-vector along the direction $\theta$.

The $k$-th element of ULA at the UE  is assumed to be located at  $\UEk{k} = \mb{u} + (k-1)\du\ori$, where  $\du$ is the inter-antenna spacing and $\ori = \left[0,1,0\right]^T$. The RIS is assumed to be located at $\mathbf{s}$ in  $xy$ plane  and have $\nris = \nx \ny$ elements arranged as a uniform planar array (UPA) as shown in Fig. \ref{fig:system}, where $\nx$ and $\ny$ represent the number of elements along $x$ and $y$ axes, respectively. Let $\dx$ and $\dy$ denote the RIS inter-element spacings along the $x$ and $y$ axes, respectively. Thus, location of the $n$-th element becomes $\sn = \mb{s} + \left[(n_x-1)\dx,~(n_y-1)\dy,~0\right]^T$,  such that $n\triangleq (n_x-1)\ny + n_y$  for $n_x=1,\dots,\nx$ and $n_y=1,\dots,\ny$. 

For the anchor-free localization, it is assumed that the UE transmits reference pulses and localizes itself using received signal reflected off the RIS. 
Similar to \cite{Kamran_2022,Kim_2024,Kim_2025_FMCW}, we assume  a full duplex UE   that can simultaneously transmit and receive the signal to perform anchor-free localization.
Let $\mb{S}\in\mathbb{C}^{K\times L}$ be reference sequence matrix with orthonormal rows transmitted over $L$ slots such that $\mb{S} \mb{S}^H = \tfrac{P_T}{K } \mb{I}_K$, where $P_T$ is the transmission power.  By the virtue of  channel reciprocity, the channel response of RIS-UE link is transpose of UE-RIS link. 
Thus, the signal received at the UE can be written as
\begin{equation}
    \mb{Y} = \mb{A}^T \operatorname{diag}\left(\mb{v}\right) \mb{A} \mb{S} + \mb{W},\label{eq:rec_signal}
\end{equation}
where $\mb{A}\in\mathbb{C}^{N\times K}$ is the channel response of UE-RIS link, $\mb{W}\in\mathbb{C}^{K\times L}$  is noise matrix comprising i.i.d. zero mean complex Gaussian entries of variance $\sigma^2$ and $\mb{v} = \left[v_1,\dots,v_n,\dots,v_N\right]^T$ is the RIS phase shift vector. Here, $v_n = e^{j\omega_n}$ with $\omega_n \in [0,2\pi]$. 
 Note that  the UE is assumed to be located in the near-field of RIS with dominant LoS component.
 Let  $\mb{A}=\left[\mb{a}_{1}, \dots, \mb{a}_{k}, \dots, \mb{a}_{K}\right]$ such that  $\mb{a}_k$ is 
 the channel between $\mb{u}_k$
 and the RIS as 
\begin{equation}
    \mb{a}_{k} =  \left[e^{-j\frac{2\pi}{\lambda}( r_{k,1}-r)},\dots,e^{-j\frac{2\pi}{\lambda}( r_{k,N}-r)}\right]^T,\label{eq:a_k}
\end{equation}
where $r_{k,n} = \norm{\UEk{k} - \sn}$ denotes the distance between between $\UEk{k}$ and $\sn$. Applying Fresnel approximation as stated in \cite{Liu_2023,srikar_2025} yields 
\begin{align*}
    r_{k,n} &= r + \frac{\left(k-1\right)^2\du^2 + \norm{\sn}^2}{2r} \\
    &+ (k-1)\du\left(\evec^T\ori - \frac{\ori^T\sn}{r}\right) -\evec^T\sn.
\end{align*}

%% file: 3_RIS_assisted_anchor_free.tex
This section  presents the proposed RIS-assisted anchor-free localization method.  
By exploiting the ability of RIS to create smart propagation environment, we design $P$ configurations that can be pre-defined for a specific AoI.
We aim to design the RIS configurations that can ensure  better signal reception across AoI.
Next, we present a two-stage low complexity localization algorithm that utilizes the reflected signals from the RIS under optimally pre-configured phase shift profiles. 

\subsection{Optimal RIS Configurations}
\label{sec:optimal_RIS_config}
In this section, we preset the design optimal RIS phase shift vector for each block in AoI that consists of $M$ grid points. For this, we aim to maximize the trace of FIM averaged over the block, as it will help to lower the   lower bound of error variance.  
Let $r_{m,p}$ and $\theta_{m,p}$ correspond to the distance and direction parameters of $m$-th grid point of $p$-th block. We refer  $m$-th grid point of $p$-th block as $(m,p)$-th point for notational simplicity. 
The reflected signal received at $(m,p)$-th point is
\begin{equation*}
    \mb{Y}_{m,p} = \mb{A}_{m,p}^T \operatorname{diag}\left(\mb{v}_p\right) \mb{A}_{m,p} \mb{S} + \mb{W}_{m,p},
\end{equation*}
such that $\mathbf{v}_p$ is the RIS phase shift vector for the $p$-th block and $\mb{A}_{m,p}$ is the channel matrix as observed by $(m,p)$-th point.
For brevity, we omit the subscript of noise matrix $\mb{W}_{m,p}$ as its entries are independent across ULA elements and transmission slots. 
Now, we post-multiple  $\mb{Y}_{m,p}$ with pseudo-inverse of  reference sequence  $\mb{S}$ and vectorize it as
\begin{equation}\label{eq:z_p}
    \mb{z}_{m,p} = \operatorname{vec}\left(\mb{Y}_{m,p}\pinv{\mb{S}}\right) = \mb{H}_{m,p}\mb{v}_p + \tilde{\mb{w}},
\end{equation}
where $\mb{H}_{m,p} = \left(\mb{A}_{m,p}^T \circ \mb{A}_{m,p}^T\right)$ and $\tilde{\mb{w}} = \frac{K}{P_T}\operatorname{vec}\left(\mb{W}\mb{S}^H\right)$.
Eq. \eqref{eq:z_p} follows from the identity $\operatorname{vec}(\mb{A}\operatorname{diag}(\mb{b})\mb{C}) = (\mb{C}^T \circ \mb{A})\mb{b}$ given in \cite{Magnus_2019_matrix} and the fact that $\pinv{\mb{S}} = \frac{K}{P_T}\mb{S}^H$.
We denote $\mb{H}_{m,p}$ as the effective channel matrix.


The covariance matrix of $\tilde{\mb{w}}$ is obtained as
\begin{align}
    \mb{R} &= \mathbb{E}\left[\tilde{\mb{w}} \tilde{\mb{w}}^H\right]\nonumber\\
    &\stackrel{(a)}{=} \left(\frac{K}{P_T}\right)^2(\mb{S}^* \otimes \mb{I}_K) \sigma^2 \mb{I}_{K^2} (\mb{S}^* \otimes \mb{I}_K)^H  \notag \\
    & \stackrel{(b)}{=} \sigma^2 (\mb{S}^* \mb{S}^T) \otimes (\mb{I}_K) \notag \\
    & \stackrel{(c)}{=} \frac{K\sigma^2}{P_T} \mb{I}_{K^2}, \label{eq:cov_mat}
\end{align}
where step (a) follows from the identity $\operatorname{vec}(\mb{A}\mb{B}) = (\mb{B}^T \otimes \mb{I}_n) \operatorname{vec}(\mb{A})$ \cite{Magnus_2019_matrix}, 
 step (b) follows from the properties of Kronecker product, and 
 step (c) follows since $(\mb{S}\mb{S}^H)^* = \tfrac{P_T}{K} \mb{I}_{K}$.

Let $\bm{\psi} = [r,\theta]^T$ be the parameter vector.  
The log-likelihood function of $\mb{z}_{m,p}$ is
\begin{equation*}
    \ln p\left(\mb{z}_{m,p};\mb{H}_{m,p}\left(\bm{\psi}\right)\right) \propto - \frac{P_T}{K\sigma^2}\norm{\mb{z}_{m,p}-\mb{H}_{m,p}\left(\bm{\psi}\right)\mb{v}_p}^2. \notag
\end{equation*}
Thus, the $(i,j)$-th element of FIM  can be obtained as
\begin{align}
    \mathcal{F}(\bm{\psi})_{i,j} &= - \mathbb{E}\left[ \pdv{\ln p(\mb{z}_{m,p};\mb{H}_{m,p}\left(\bm{\psi}\right))}{\psi_i}{\psi_j} \right],\notag\\
    &= \frac{2P_T}{K\sigma^2}\operatorname{Re}\left\{\mb{v}_p^H \pdv{\mb{H}_{m,p}\left(\bm{\psi}\right)^{H}}{\psi_i} \pdv{\mb{H}_{m,p}\left(\bm{\psi}\right)}{\psi_j}\mb{v}_p\right\}, \notag
\end{align}
where $\operatorname{Re}\{\cdot\}$ denotes the real part of a complex number.
Thus, the trace of FIM $\mathcal{F}(\bm{\psi})$ observed at $(m,p)$-th point becomes
\begin{align}\label{eq:trace_FIM}
    \mb{Q}_{m,p} = \frac{2P_T}{K\sigma^2}\operatorname{Re}\left\{\mb{v}_p^H \mb{G}_{m,p} \mb{v}_p\right\},
\end{align}
where $\mb{G}_{m,p} = \pdv{\mb{H}_{m,p}\left(\bm{\psi}\right)^{H}}{r} \pdv{\mb{H}_{m,p}\left(\bm{\psi}\right)}{r} + \pdv{\mb{H}_{m,p}\left(\bm{\psi}\right)^{H}}{\theta} \pdv{\mb{H}_{m,p}\left(\bm{\psi}\right)}{\theta}$. The partial differentiation in $\mb{G}_{m,p}$ can be evaluated as given below. 
Using \eqref{eq:a_k}, we have the $(i(q,l),n)$-th element of $\mathbf{H}$ as 
\begin{equation*}
    \mb{H}_{i(q,l),n}^t = \mb{A}_{n,t} \mb{A}_{n,q} \triangleq \exp(\frac{-j2\pi}{\lambda}\mb{B}_{i(q,l),n}^t),
\end{equation*}
where 
\begin{align*}
    \mb{B}_{i(q,l),n}^t &=  \frac{\norm{\sn}^2}{r} -2\sn^T\evec + \frac{\Delta \du^2}{2r} + \delta \du\left(\ori^T\evec - \frac{\ori^T\sn}{r}\right),
\end{align*}
for $n=1,\dots,N$, $t,l,q=1,\dots,K$ and $i(q,l) \triangleq K(l-1)+q$ such that $\delta = (t+q)-2$ and $\Delta = (t-1)^2 + (q-1)^2$.
Observe that the partial derivatives of $\mathbf{H}$ with respect to $r$ and $\theta$ can be obtained as
\begin{align*}
    \pdv{\mb{H}}{r} = \frac{-j2\pi}{\lambda}\mb{H} \odot \pdv{\mb{B}}{r}, \text{~and~} \pdv{\mb{H}}{\theta} = \frac{-j2\pi}{\lambda}\mb{H} \odot \pdv{\mb{B}}{\theta}.
\end{align*}
The partial derivatives of $\mb{B}_{i(q,l),n}$ are obtained as
\begin{align*}
    \pdv{\mb{B}_{i(q,l),n}^t}{r} &= \frac{\delta \du \ori^T\sn - \norm{\sn}^2 - \Delta \du^2/2 }{r^2}, \text{~and}\\
    \pdv{\mb{B}_{i(q,l),n}^t}{\theta} &= \left(\delta \du g^T - 2\sn^T\right) \pdv{\evec}{\theta},
\end{align*}
where $\pdv{\evec}{\theta} = \left[ -\sin{\theta},~\cos{\theta},~0 \right]^T$.

Next, we design the $p$-th configuration of RIS to maximize the  trace of FIM averaged over the $p$-th block, i.e. 
\begin{subequations}\label{eq:opt_FIM_full}
    \begin{align}
        \underset{\mb{v}_p}{\max} \quad & \frac{1}{M} \sum_{m = 1}^{M} \mb{Q}_{m,p}, \label{eq:opt_FIM_full_obj} \\
        \text{s.t.} \quad & |\mb{v}_{p_n}| = 1 \quad \forall ~ n=1,\dots,N. \label{eq:opt_FIM_full_cond}
    \end{align}
\end{subequations}
The above optimization problem can be reformulated as 
\begin{subequations}\label{eq:RayleighQuo}
    \begin{align}
        \underset{\mb{v}_p}{\max} \quad & \mb{v}_p^H \mb{G}_{p} \mb{v}_p, \label{eq:RayleighQuo_obj} \\
        \text{s.t.} \quad & |\mb{v}_{p_n}| = 1 \quad \forall ~ n=1,\dots,N, \label{eq:RayleighQuo_cond}
    \end{align}
\end{subequations}
where $\mb{G}_{p} = \sum_{m = 1}^{M} \mb{G}_{m,p}$.
This is a  Rayleigh quotient maximization problem and its  solution can be obtained using the dominant eigenvector of $\mb{G}_{p}$ denoted using $\bm{\eta}$. Thus, enforcing  the constraint \eqref{eq:RayleighQuo_cond}, the near-optimal RIS phase shift vector for the $p$-th block can be obtained as
\begin{equation}\label{eq:RIS_opt_p}
    \mb{v}_p = \exp \left(j\angle \bm{\eta}\right).
\end{equation}
Solving \eqref{eq:RayleighQuo} for each block, we obtain the near-optimal RIS configuration matrix $\mb{V} = \left[\mb{v}_1,\dots,\mb{v}_p,\dots,\mb{v}_P\right]$.


\subsection{Anchor-free Localization}\label{sec:anchor_free_localization}
In this section, we develop a two-stage low complexity anchor-free localization algorithm. In the first stage, we obtain a coarse estimate of the UE location. The coarse estimate is considered to be a grid point nearest to the UE location. In the next stage, we apply iterative refinement to the coarse estimates to obtain off-grid UE location.


Using \eqref{eq:rec_signal} and \eqref{eq:z_p}, the vectorized received signal under $p$-th RIS configuration can be written as
\begin{equation}
    \mathbf{z}_p={\rm vec}\left(\mathbf{Y}\pinv{\mathbf{S}}\right) = \mathbf{H}\mathbf{v}_p + \tilde{\mathbf{w}}.\notag
\end{equation}
Stacking observation vectors $\mb{z}_p$ across columns gives
\begin{equation}\label{eq:obs_mat}
    \mb{Z} = \mathbf{H}\mb{V} + \tilde{\mb{W}},
\end{equation}
where $\tilde{\mb{W}}_{k,p} \sim \mathcal{CN}(0,\tfrac{K\sigma^2}{P_T})$.
The likelihood function of $\mb{Z}$ is
\begin{equation}
    \ln p\left(\mb{Z};\mb{H}\right) \propto - \frac{P_T}{K\sigma^2}\norm{\mb{Z}-\mb{H}\mb{V}}^2_F. \notag
\end{equation}
The coarse estimates  $(\check{r},\check{\theta})$ is a point in grid $\mathcal{A}$ (shown in Fig. \ref{fig:aoi}) that maximizes the likelihood function, i.e.
\begin{equation}
    (\check{r},\check{\theta}) = \underset{({r}_{m},{\theta}_{m})\in\mathcal{A}}{\operatorname{argmin}} \quad  \norm{\mb{Z} - \mb{H}(r_m,\theta_m)\mb{V}}^2_F.
\end{equation}
The above objective function can be written as
\begin{align*}
    \norm{\mb{Z} - \mb{H}\mb{V}}^2_F &= \operatorname{tr}\left( \mb{Z}^H\mb{Z} + (\mb{H}\mb{V})^H\mb{H}\mb{V}\right)\\
    & - \operatorname{tr}\left(\mb{Z}^H\mb{H}\mb{V} + (\mb{H}\mb{V})^H\mb{Z}\right).
\end{align*}
The term $(\mb{H}\mb{V})^H\mb{H}\mb{V}$ is observation independent and might lead to model mismatch due to the non-linearities in the near-field channel, thus we ignore it. The objective function can be reformulated as the  Frobenius inner product of $\mb{Z}$ and $\mb{H}\mb{V}$. Therefore, with normalization, we can reformulate the problem as  maximizing the cosine similarity defined as
\begin{equation}\label{eq:cosine_sim}
    \alpha(r,\theta) = \frac{|\operatorname{tr}\left(\mb{Z}^H\mb{H}(r,\theta)\mb{V} \right)|^2}{\norm{\mb{Z}}^2_F \norm{\mb{H}(r,\theta)\mb{V}}^2_F}.
\end{equation}
Cosine similarity is robust to noise since it prioritizes phase alignment over magnitude.
Thus, the coarse estimates $(\check{r},\check{\theta})$ can be obtained as a point in grid $\mathcal{A}$  that  maximizes  \eqref{eq:cosine_sim}.
However, to facilitate robust localization, we utilize  coarse estimates that exhibits top-$\kappa$ cosine similarity scores, i.e.
\begin{equation}\label{eq:corr}
    \Psi = \underset{({r}_{m},{\theta}_{m})\in\mathcal{A}}{{}^{\kappa}\operatorname{argmax}} \quad \alpha(r_m,\theta_m).
\end{equation}
where the operator ${}^{\kappa}\operatorname{argmax}(\cdot)$  determines the top-$\kappa$ local maxima points.
Thus, we have $\Psi = \{(\check{r}_{i},\check{\theta}_{i}): \alpha(\check{r}_{i},\check{\theta}_{i})\geq \alpha(\check{r}_{i+l},\check{\theta}_{i+l})~\forall i=1,\dots,\kappa, \text{~and~}\forall l>0\}$ denotes the set of grid points having top-$\kappa$  similarity scores.
Next, we  refine each coarse estimate by again evaluating the cosine similarity over the fine-resolution grid $\mathcal{B}_i$ defined around $(\check{r}_i,\check{\theta}_i)$ as
$$\mathcal{B}_i=[\check{r}_i - \tfrac{d_r}{2},\check{r}_i + \tfrac{d_r}{2}] \times [\check{\theta}_i - \tfrac{d_{\theta}}{2},\check{\theta}_i + \tfrac{d_{\theta}}{2}],$$
where $d_r$ and $d_{\theta}$ represent the fine grid size for $r$ and $\theta$ respectively.
Thus, the refined estimate can be obtained as
\begin{equation}\label{eq:final_est}
   (\hat{r}_i,\hat{\theta}_i) = \underset{(r_l,\theta_l) \in \mathcal{B}_i}{\operatorname{argmax}} \quad \alpha(r_l,\theta_l).
\end{equation}
Finally, we select the final estimates $(\hat{r},\hat{\theta})$ from set $\{(\hat{r}_i,\hat{\theta}_i)\}_{i=1}^\kappa$ that exhibits the highest score. 
The proposed two stage  anchor-free localization algorithm is summarized in Algorithm \ref{Alg:afloc}.
\begin{algorithm}[t!]
    \KwInput{$\mb{Z}$, $\mb{V}$, $\mathcal{A}$ and $\kappa$}
    \KwOutput{$\hat{r}, \hat{\theta}$}
    Compute cosine similarity $\alpha(r_m,\theta_m)$ for all points in grid $\mathcal{A}$ \\
    Select coarse estimate set $\Psi$ using the top-$\kappa$- similarity scores, i.e.
    {\small \begin{equation*}
        \Psi = \{(\check{r}_{i},\check{\theta}_{i}): \alpha(\check{r}_{i},\check{\theta}_{i})\geq \alpha(\check{r}_{i+l},\check{\theta}_{i+l})~\forall i=1,\dots,\kappa\}.
    \end{equation*}}
    \For{$i=1,\dots,\kappa$}{
    Design fine resolution grid for $i$-th coarse estimate
    {\small $$\mathcal{B}_i=[\check{r}_i - d_r,\check{r}_i + d_r] \times [\check{\theta}_i - d_{\theta},\check{\theta}_i + d_{\theta}]$$}\\
    Refine the $i$-th coarse estimate as
    {\small \begin{equation*}
        (\hat{r}_i,\hat{\theta}_i) = \underset{(r_l,\theta_l) \in \mathcal{B}_i}{\operatorname{argmax}} \quad \alpha(r_l,\theta_l).
    \end{equation*}}
    }
    Select the optimal estimate as
    {\small \begin{equation*}
       (\hat{r},\hat{\theta}) =  \underset{(\hat{r}_i,\hat{\theta}_i)}{\operatorname{argmax}} \quad \alpha(\hat{r}_i,\hat{\theta}_i).
    \end{equation*}}
    \caption{RIS-assisted Anchor-free Localization}
    \label{Alg:afloc}
\end{algorithm}
\setlength{\textfloatsep}{6pt}

%% file: 4_Numerical_Results.tex
\begin{figure*}[t]
  \centering
  \subfloat[Random RIS Phase Shifts \label{fig:FIM_rand}]{\includegraphics[width=0.38\textwidth]{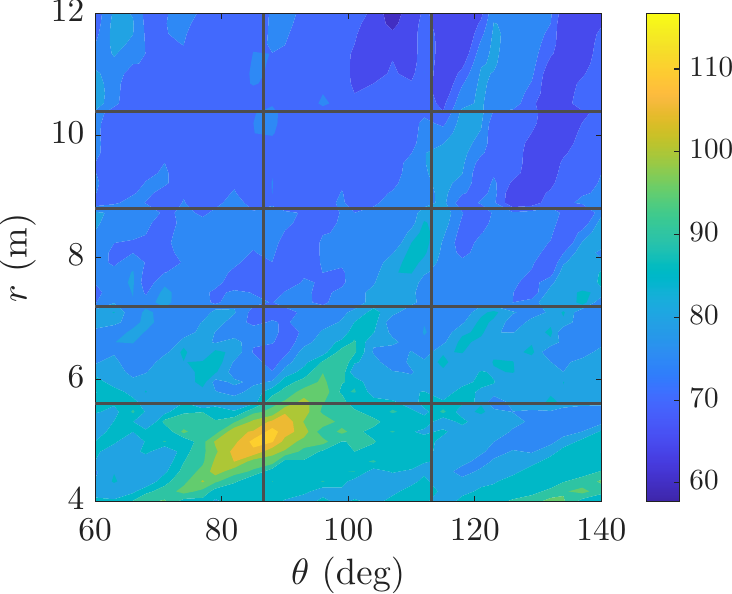}}
  \hspace{1.5cm}
  \subfloat[Optimal RIS Phase Shifts \label{fig:FIM_opt}]{\includegraphics[width=0.38\textwidth]{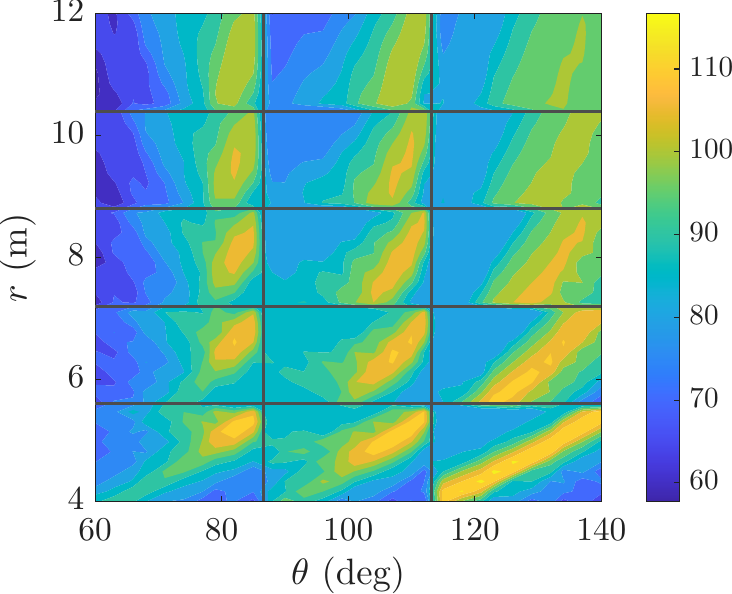}}
  \caption{Contour plot of trace of FIM across the AoI for $N=196$, $K=12$, $P_r = 5$, $P_{\theta} = 3$ and grid $\mathcal{G}_2$. }
  \label{fig:FIM}
\end{figure*}

\begin{figure*}[t]
  \centering
  \subfloat[Distance $r$ \label{fig:snr_r}]{\includegraphics[width=0.38\textwidth]{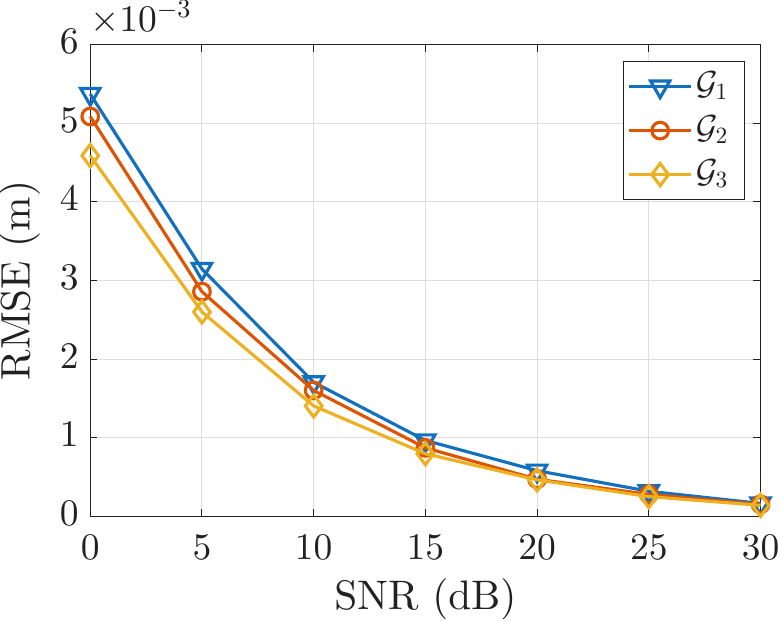}}
  \hspace{1.5cm}
  \subfloat[Direction $\theta$ \label{fig:snr_theta}]{\includegraphics[width=0.38\textwidth]{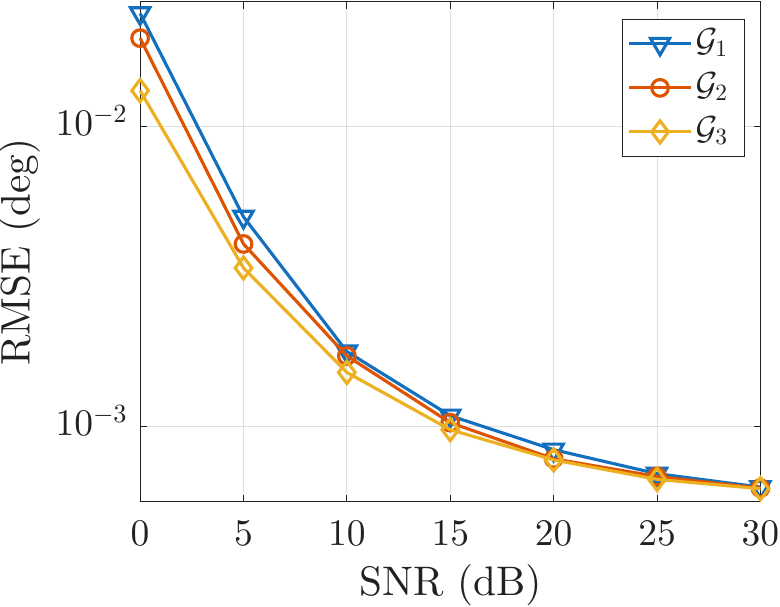}}
  \caption{RMSE vs. SNR for $N = 196$, $K=12$, $P_r = 5$ and $P_{\theta} = 3$ with $\mathcal{G}_1 = (12 \times 8)$, $\mathcal{G}_2 = (10 \times 10)$ and $\mathcal{G}_3 = (16 \times 12$).}
  \label{fig:rmse_v_SNR}
\end{figure*}

This section presents the numerical analysis of the RMSE performance of Algorithm \ref{Alg:afloc}. For the numerical results, we considered the parameters as location of RIS $\mathbf{s} = [5,5,5]^T$, number of RIS elements $N\in\{144,256,324\}$ with $\nx = \ny = \sqrt{\nris}$, operational wavelength $\lambda = 0.1$ m, inter-antenna element spacings $\du = \dx = \dy = \frac{\lambda}{2}$, transmission power $P_T = 30$ dBm,  number of UE antennas $K \in [8,20]$, number of transmission slots $L=K$; unless mentioned otherwise.   
We define the near-field region  between Fresnel distance $0.62 \lambda^{-1/2} \left(a_R^2+b_R^2\right)^{3 / 4}$ and Rayleigh distance $2\lambda^{-1}\left(a_R^2+b_R^2\right)$ \cite{srikar_2025,Liu_2023}, where $a_R = (\nx - 1)\dx$ and $b_R = (\ny - 1)\dy$ denote the dimensions of the RIS along $y$ and $z$ axes. 
The AoI is defined for distance $[4,12]$m and direction $[60^{\circ},140^{\circ}]$. Number of AoI blocks across distance and direction are $P_r = 5$ and $P_{\theta} = 3$ respectively. For each AoI block, we considered three block sizes ($M_r \times M_{\theta}$) as $\mathcal{G}_1: (12 \times 8)$, $\mathcal{G}_2: (10 \times 10)$ and $\mathcal{G}_3: (16 \times 12)$. The number of coarse estimates $\kappa = 5$ for local refinement. The fine grid sizes for $r$ and $\theta$ are $d_r = 1$ m and $d_{\theta} = 20^{\circ}$, respectively.
We perform the Monte Carlo simulation with  10,000  iterations to evaluate the RMSE performance of the proposed algorithm. In each iteration, we sample the UE location $(r,\theta)$ uniformly at random in the AoI.

Fig. \ref{fig:FIM} shows the trace of FIM (see \eqref{eq:trace_FIM}) across the AoI  for both random RIS phase shift and optimal RIS phase shift configurations proposed in Section \ref{sec:phase_design}.
Fig. \ref{fig:FIM_rand} shows that the random configurations exhibit a higher peak value for a very small region in the AoI. However, the remaining region exhibits significantly lower values of trace of FIM, which implies that the error variance will be large in this region. 
In contrast, Fig. \ref{fig:FIM_opt} shows that the proposed optimization strategy exhibits broadly similar Fisher information characteristics for each AoI block, which in turn will ensure consistently lower  error variance across AoI. 

\begin{figure*}[t]
  \centering
  \subfloat[Distance $r$ \label{fig:NK_r}]{\includegraphics[width=0.38\textwidth]{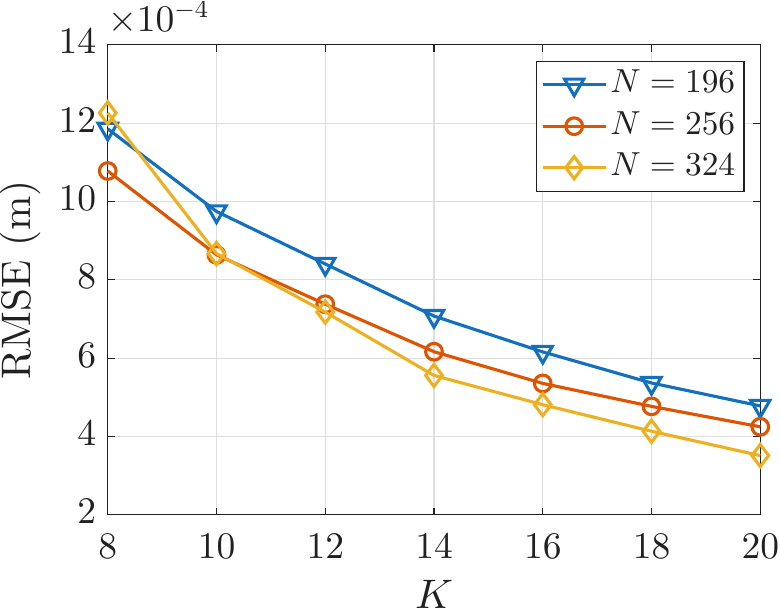}}
  \hspace{1.5cm}
  \subfloat[Direction $\theta$ \label{fig:NK_theta}]{\includegraphics[width=0.38\textwidth]{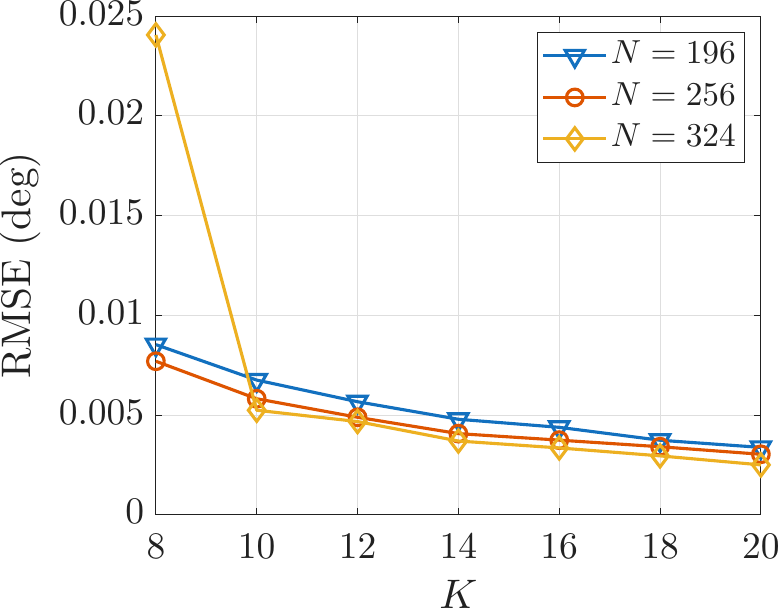}}
  \caption{RMSE vs. K for SNR = $15$dB, $P_r = 5$, $P_{\theta} = 3$ and grid $\mathcal{G}_2 = (10 \times 10)$.}
  \label{fig:rmse_v_NK}
\end{figure*}

Fig. \ref{fig:rmse_v_SNR} demonstrates that the RMSE performance of both the parameters improve with increase in SNR, as expected. Further,  it can be seen that the  larger block size $\mathcal{G}$ used for coarse estimation improves the RMSE, however this gain diminishes with the increase in SNR. 
Furthermore, it is observed that the distance estimate is relatively more sensitive to the overall grid points over the block size. Fig. \ref{fig:snr_r} shows that $\mathcal{G}_2$ achieves lower RMSE than $\mathcal{G}_1$ for $r$ despite it having lower resolution in distance. 
This is attributed to $\mathcal{G}_2$ having more number of grid points $(10\times 10\times 5 \times 3)$ across AoI as compared to $(12\times 8 \times 5 \times 3)$ grid points in $\mathcal{G}_1$.
Thus, the location RMSE performance improves with increase in the grid resolution.

The effect of $N$ and $K$ on the RMSE performance is demonstrated in Fig. \ref{fig:rmse_v_NK}. For both $r$ and $\theta$, the RMSE performance is observed to improve with  increase in either $N$ or $K$. However, the performance improvement is relatively larger for $r$ (see Fig. \ref{fig:NK_r}) as compared to the improvements observed for $\theta$ (see Fig. \ref{fig:NK_theta}). This is because $r$ is more sensitive to changes in $N$ and $K$, as is also evident from the partial derivatives in given Subsection \ref{sec:optimal_RIS_config}. In summary, it is safe to say that the  localization performance improves with the increase in RIS and UE's ULA dimensions.

%% file: 5_Conclusion.tex
This paper first presents optimal RIS configurations to maximize block-wise averaged trace of FIM across the AoI in order to lower the localization error variance. 
Next, a two-stage RIS-assisted anchor-free localization algorithm is presented that first obtains the coarse estimates by evaluating the cosine symmetry between a coarse grid and the received signal observed under the pre-defined optimally configured RIS. The second stage involves fine-grid refinement of coarse estimates using a similar approach. Through extensive numerical analysis, we demonstrated that the proposed algorithm achieves low RMSE performance for practical values of SNR. 

%% file: main.bbl

%% file: main.bbl
\begin{thebibliography}{10}
\providecommand{\url}[1]{#1}
\csname url@samestyle\endcsname
\providecommand{\newblock}{\relax}
\providecommand{\bibinfo}[2]{#2}
\providecommand{\BIBentrySTDinterwordspacing}{\spaceskip=0pt\relax}
\providecommand{\BIBentryALTinterwordstretchfactor}{4}
\providecommand{\BIBentryALTinterwordspacing}{\spaceskip=\fontdimen2\font plus
\BIBentryALTinterwordstretchfactor\fontdimen3\font minus \fontdimen4\font\relax}
\providecommand{\BIBforeignlanguage}[2]{{%
\expandafter\ifx\csname l@#1\endcsname\relax
\typeout{** WARNING: IEEEtran.bst: No hyphenation pattern has been}%
\typeout{** loaded for the language `#1'. Using the pattern for}%
\typeout{** the default language instead.}%
\else
\language=\csname l@#1\endcsname
\fi
#2}}
\providecommand{\BIBdecl}{\relax}
\BIBdecl

\bibitem{Kamran_2023}
K.~Keykhosravi and et~al., ``Leveraging {RIS}-enabled smart signal propagation for solving infeasible localization problems: Scenarios, key research directions, and open challenges,'' \emph{IEEE Vehicular Technology Magazine}, vol.~18, no.~2, pp. 20--28, 2023.

\bibitem{Zhao_6g_white_paper_near_field}
\BIBentryALTinterwordspacing
Y.~Zhao and et~al., ``{6G} near-field technologies white paper 2.0,'' 2025. [Online]. Available: \url{https://eprints.gla.ac.uk/354426/}
\BIBentrySTDinterwordspacing

\bibitem{bourdoux20206gwhitepaperlocalization}
\BIBentryALTinterwordspacing
A.~Bourdoux and et~al., ``{6G} white paper on localization and sensing,'' 2020. [Online]. Available: \url{https://arxiv.org/abs/2006.01779}
\BIBentrySTDinterwordspacing

\bibitem{srikar_2025}
\BIBentryALTinterwordspacing
S.~S. Sadhu, P.~D. Mankar, and S.~Nannuru, ``Near-field {5D} pose estimation using reconfigurable intelligent surfaces,'' in \emph{IEEE Global Communications Conference (GLOBECOM)}, 2025. [Online]. Available: \url{https://arxiv.org/abs/2505.01829}
\BIBentrySTDinterwordspacing

\bibitem{Sarah_2021}
S.~Basharat and et~al., ``Reconfigurable intelligent surfaces: Potentials, applications, and challenges for {6G} wireless networks,'' \emph{IEEE Wireless Communications}, vol.~28, no.~6, pp. 184--191, 2021.

\bibitem{1gto5g_2018}
J.~A. del Peral-Rosado and et~al., ``Survey of cellular mobile radio localization methods: From {1G} to {5G},'' \emph{IEEE Communications Surveys \& Tutorials}, vol.~20, no.~2, pp. 1124--1148, 2018.

\bibitem{Cheng_2012_survey}
L.~Cheng, C.~Wu, Y.~Zhang, H.~Wu, M.~Li, and C.~Maple, ``A survey of localization in wireless sensor network,'' \emph{International Journal of Distributed Sensor Networks}, vol.~8, no.~12, p. 962523, 2012.

\bibitem{Zafari_2019}
F.~Zafari, A.~Gkelias, and K.~K. Leung, ``A survey of indoor localization systems and technologies,'' \emph{IEEE Communications Surveys \& Tutorials}, vol.~21, no.~3, pp. 2568--2599, 2019.

\bibitem{Tapas_2018}
T.~K. Mishra and et~al., ``An enhanced path planning model for anchor-free localization in wireless sensor networks,'' in \emph{International Conference on Information Technology (ICIT)}, 2018, pp. 204--209.

\bibitem{Sekhar_2021}
V.~C. S.~R. Rayavarapu and A.~Mahapatro, ``A novel range-free anchor-free localization in {WSN} using sun flower optimization algorithm,'' in \emph{2021 Advanced Communication Technologies and Signal Processing (ACTS)}, 2021, pp. 1--6.

\bibitem{Premachandra_2023}
H.~A. G.~C. Premachandra and et~al., ``{UWB} radar {SLAM}: An anchorless approach in vision denied indoor environments,'' \emph{IEEE Robotics and Automation Letters}, vol.~8, no.~9, pp. 5299--5306, 2023.

\bibitem{Cai_2024}
X.~Cai, L.~Wang, S.~Sun, H.~Han, and K.~Han, ``Anchor-free relative {3D} pose estimation using ultra-wideband and inertial data fusion,'' in \emph{2024 IEEE SENSORS}, 2024, pp. 1--4.

\bibitem{Chen_2023}
H.~Chen, H.~Kim, M.~Ammous, G.~Seco-Granados, G.~C. Alexandropoulos, S.~Valaee, and H.~Wymeersch, ``{RIS}s and sidelink communications in smart cities: The key to seamless localization and sensing,'' \emph{IEEE Communications Magazine}, vol.~61, no.~8, pp. 140--146, 2023.

\bibitem{Kamran_2022}
K.~Keykhosravi, G.~Seco-Granados, G.~C. Alexandropoulos, and H.~Wymeersch, ``{RIS}-enabled self-localization: Leveraging controllable reflections with zero access points,'' in \emph{ICC 2022 - IEEE International Conference on Communications}, 2022, pp. 2852--2857.

\bibitem{Kim_2024}
H.~Kim, H.~Chen, M.~F. Keskin, Y.~Ge, K.~Keykhosravi, G.~C. Alexandropoulos, S.~Kim, and H.~Wymeersch, ``{RIS}-enabled and access-point-free simultaneous radio localization and mapping,'' \emph{IEEE Transactions on Wireless Communications}, vol.~23, no.~4, pp. 3344--3360, 2024.

\bibitem{Mustafa_2025}
M.~Ammous and et~al., ``{3D} cooperative positioning via {RIS} and sidelink communications with zero access points,'' \emph{IEEE Transactions on Mobile Computing}, vol.~24, no.~7, pp. 6119--6136, 2025.

\bibitem{Kim_2025_FMCW}
\BIBentryALTinterwordspacing
N.~Hyowon~Kim, M.~F. Keskin, Z.~S. He, J.~Gil, G.-S. Granados, and H.~Wymeersch, ``{RIS}-enabled self-localization with {FMCW} radar,'' 2025. [Online]. Available: \url{https://arxiv.org/abs/2503.21021}
\BIBentrySTDinterwordspacing

\bibitem{Liu_2023}
Y.~Liu, Z.~Wang, J.~Xu, C.~Ouyang, X.~Mu, and R.~Schober, ``Near-field communications: A tutorial review,'' \emph{IEEE Open Journal of the Communications Society}, vol.~4, pp. 1999--2049, 2023.

\bibitem{Magnus_2019_matrix}
J.~R. Magnus and H.~Neudecker, \emph{Matrix Differential Calculus with Applications in Statistics and Econometrics}.\hskip 1em plus 0.5em minus 0.4em\relax John Wiley \& Sons, Inc., 2019, ch.~3, pp. 34--36.

\end{thebibliography}
